\UseRawInputEncoding

\documentclass[preprint, 12pt, a4paper]{elsarticle}

\usepackage{hyperref}

\usepackage{amssymb}

\usepackage{lineno}

\journal{ArXiv}

\def\x{$\times$}

\begin{document}

\begin{frontmatter}



\title{Counting Molecules: Python based scheme for automated enumeration and categorization of molecules in scanning tunneling microscopy images}


\author[FZU,Olomouc]{Jack Hellerstedt}
\ead{hellerstedt.jack@gmail.com}
\author[FZU,Olomouc]{Ale\v{s} Cahl\'{i}k}
\author[FZU,Olomouc]{Martin \v{S}vec}
\author[FZU]{Oleksandr Stetsovych}
\author[Aachen]{Tyler Hennen}

\address[FZU]{Institute of Physics of the Czech Academy of Sciences, v.v.i., Cukrovarnick\'{a} 10, 162 00 Praha 6, Czech Republic}
\address[Olomouc]{Regional Centre of Advanced Technologies and Materials, Palack\'{y} University, \v{S}lechtitel\r{u} 27, 78371 Olomouc, Czech Republic}
\address[Aachen]{Institut f\"{u}r Werkstoffe der Elektrotechnik II, RWTH Aachen University, Sommerfeldstra{\ss}e 24, 52056
Aachen, Germany}

\begin{abstract}
Scanning tunneling and atomic force microscopies (STM/nc-AFM) are rapidly progressing to offer unprecedented spatial resolution of a diverse array of chemical species.  In particular, they are employed to characterize on-surface chemical reactions by directly examining precursors and products.  Chiral effects and self-assembled structures can also be investigated.  This open source, modular, python based scheme automates the categorization of a variety of molecules present in medium sized (10$\times$10 to 100$\times$100 nm) scanned probe images.

\end{abstract}


\end{frontmatter}


\section{Motivation and significance}

Scanned probe techniques such as scanning tunneling microscopy (STM) and non-contact atomic force microscopy (nc-AFM) have now made direct investigation of on-surface reactions a routine experimental technique \cite{Barth2007, Jelinek2017}. As the diverse ecosystem of organic molecules accessible by these techniques continues to grow so does the need for more sophisticated tools to extract quantitative information from larger and more complicated datasets of molecule imaging.

Extracting statistics from STM images is often done by hand \cite{Capsoni2016, Prinz2015, Stetsovych2016, Goll2022}.  This necessarily limits the size and complexity of statistical problems that can be tackled using these datasets.  To address this limitation in extracting and categorizing molecules in STM images, we developed an automated scheme utilizing existing image processing libraries written using the Python programming language.

The existing widely used programs for the analysis of scanned probe data (WSxM \cite{Horcas2007} and Gwyddion \cite{Necas2012}) do not presently incorporate tools for automated image processing and feature extraction.  ImageJ \cite{Rueden2017} is a powerful and widely used piece of software in the biology community, but lacks compatibility and relevance for the datasets this package is designed to address.  Another hindrance is that these image processing tools for biology applications are often written using commercial, closed source MATLAB code. Digital Surf has feature categorization tools but is also commercial and closed source \cite{Cognard2020}.

There has been recent progress in developing machine-learning based tools for the automated assignment of atomic structure and defect information from scanned probe datasets, \cite{Scherbela2017, Ziatdinov2017, Li2021} as well as automated routines for improving the probe condition in STM \cite{Rashidi2018, Krull2020}.  However these solutions address comparatively homogeneous datasets of more simple structures compared the present case, or require the computational resources and data (real or artificial) to train a convolutional neural network \cite{Krull2020, Li2021}.  We wanted to avoid this approach and instead develop a lightweight tool to quickly identify statistical trends in a large number of STM images with a diverse manifold of initially unknown molecular species; the functions offered in this package were used to prepare Fig. 3 in Hellerstedt \emph{et. al} \cite{Hellerstedt2019}.

Our approach addresses some specific obstacles.  There are practical limitations to the data resolution versus the number of species sampled.  The sorting method needs account for different adsorption configurations of the same species (\emph{e.g.}, rotations and chirality).  While explicit pairwise template matching has advantages in robustness, it suffers from the computational demands scaling in quadrature with the number of molecules to be sorted.

Our solution to these problems relies on the Zernike polynomial basis set to provide a ``fingerprint" of coefficients representing every molecule \cite{Khotanzad1990, Coelho2013}.  These coefficients are robust in response to rotations and noise in the data. In conjunction with other physically motivated coefficients, they provide the input to clustering algorithms for sorting the molecules.  There are many different algorithms that can be applied to this type of data for the purposes of grouping like kinds of molecules together.  We discuss some of the approaches we applied and their relative efficacy.

\section{Software description}

This package consists of a set of functions written using the Python libraries Numpy \cite{Oliphant2007}, scikit-image \cite{VanderWalt2014}, scikit-learn \cite{Pedregosa2011}, Mahotas \cite{Coelho2013} and Matplotlib \cite{Hunter2007}.  In practice, it is a series of functions called sequentially as shown in Fig. \ref{flowchart}.  We developed the current implementation to be useful `out-of-the-box' with only minimum skill prerequisites required (\emph{e.g.}, installing a python distribution and necessary packages).  The modular nature of the data processing flow allows for a great deal of flexibility to address different datasets.  The open source, repository based distribution of this package allows for further customization, and the possibility for community based development and improvement on these core functionalities.

\subsection{Software Architecture}

We first read in the image data and apply filters to make the image suitable for molecule template extraction using an adaptive threshold. After identifying closed contours corresponding to the perimeters of each molecule, we use the interior data of each contour as a representative template image. We then extract numerical features from each molecule, consisting of the calculated Zernike moments \cite{Khotanzad1990, Coelho2013} of the template images as well as the maximum topographic heights and the contour perimeter lengths. Finally, we use clustering algorithms \cite{Pedregosa2011} to categorize these features. Further functionality is provided to visualize the sorted categories, and make manual corrections.

\begin{figure}
\centering
\includegraphics[keepaspectratio=true, width=.8\linewidth]{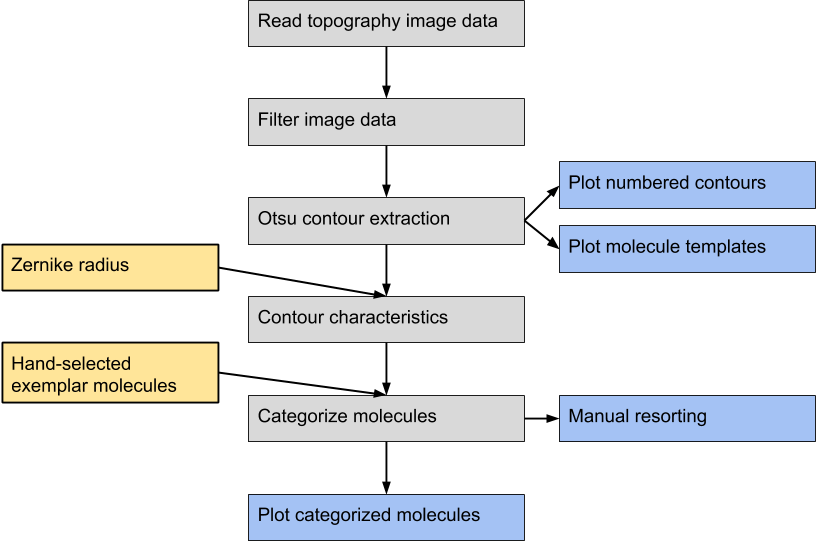}
\caption{Architecture and usage of main functions.  Yellow boxes indicate optional user input parameters, blue boxes are visual outputs of the program.}
\label{flowchart}
\end{figure}

Figure \ref{flowchart} shows the typical use of the provided functions.  We rely on the Nanonispy library to read in data from one of the prevalent formats (Nanonis SXM files).  This generates an array of pixel data and the rescaling factors to convert to real-space distances.

\subsection{Software Functionalities}

The default filtering function provided performs a Gaussian filter and plane fit subtraction of the image data.  We calculate the global Otsu threshold \cite{Otsu1979} and use that to scale the offset value for feature extraction using a local thresholding method.  We include diagnostic functions that can plot the filtered image data with the extracted contours, labeled by number, as well as a grid view of all the extracted molecule templates.

Zernike moments are the coefficients for representing an image decomposed into the orthogonal Zernike polynomial basis set \cite{Khotanzad1990}.  This method has attracted attention and effort in the computer vision community for being rotationally invariant.  This property is particularly useful in the context of our application because we wish to match molecules regardless of how they are absorbed on the sample surface.  We utilize the Mahotas library for calculating the Zernike moments for each molecule template \cite{Coelho2013}. We set the median template diagonal length as the default value of the Zernike radius input, an assumption that could fail depending on the homogeneity of the templates.

In addition to rotational invariance, these moments are insensitive to translation, mirroring, and rescaling.  To account for differences in the real space footprint of each molecule, we additionally calculate the length, as well as the maximum pixel value within each contour.

With these characteristic moments and physical length scales, we perform a clustering analysis using the algorithms available in the Scikit-learn library \cite{Pedregosa2011}.  For the datasets studied in this work, we found the Birch algorithm \cite{Zhang1996} with a threshold factor between 0.1 to 0.4 to be most effective at sorting images with no \emph{a priori} knowledge of the number of categories.  When the number of sorting categories is known, hierarchical clustering was found to be more accurate.  The most effective sorting was accomplished using affinity propagation, where the cluster center preferences were defined using a hand-selected set of exemplar molecules.

The invariance of the Zernike moments to mirror symmetry in particular, makes them insensitive to differentiating between chiral molecules.  Absolute quantifications of chirality are surprisingly difficult to define \cite{Buda1992}.  We developed a function to do a pairwise comparison of molecules within each sorted category.  By comparing each molecule and its mirror image, they can be sorted into right- and left- handed categories.

All these image, contour, correlations, and categorization data can be  exported/ saved for later use.  Also included is a function using the interactive features of Matplotlib to manually categorize all the molecules.

\section{Illustrative Examples}

\begin{figure}
\centering
\includegraphics[keepaspectratio=true, width=1\linewidth]{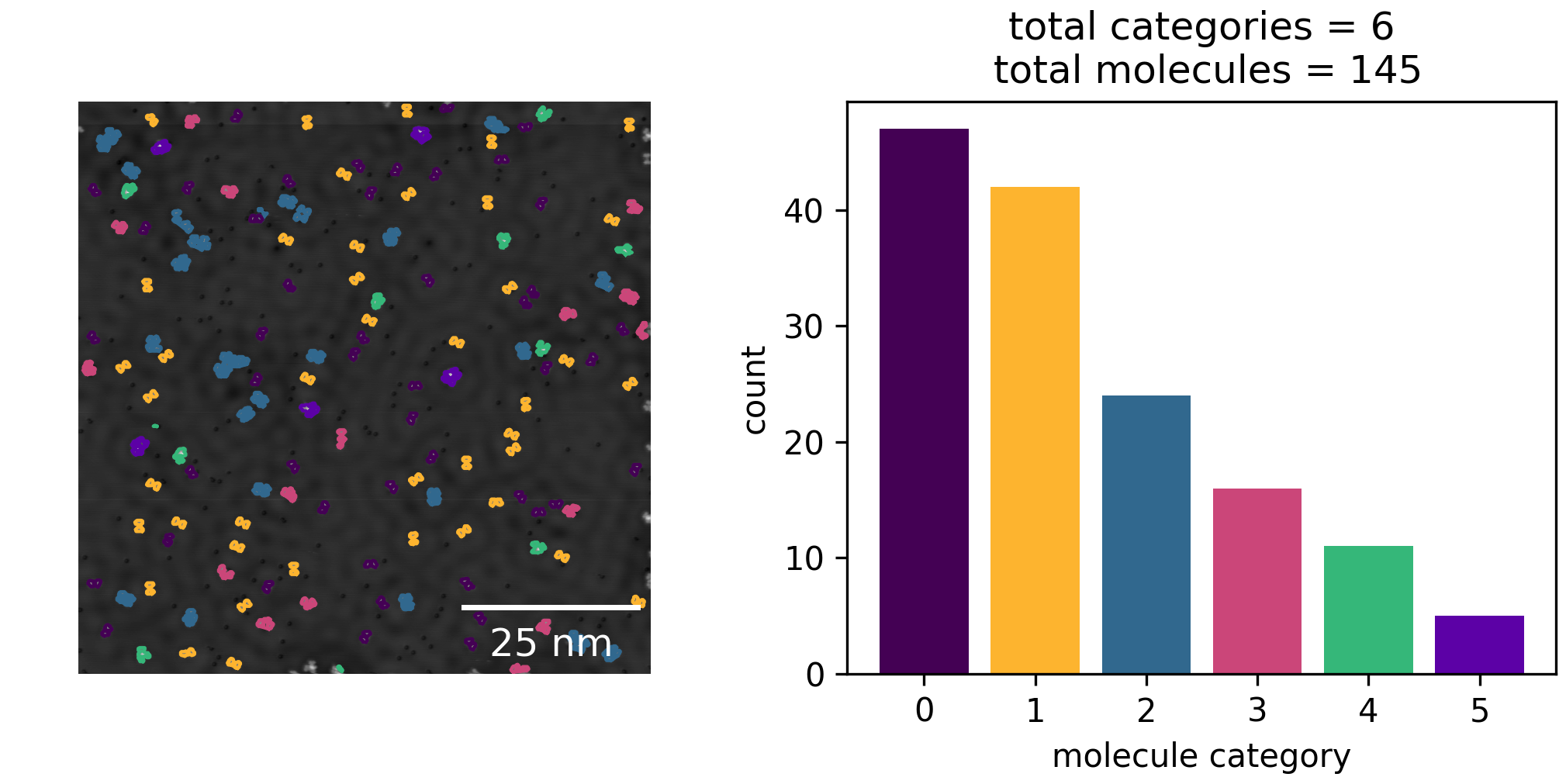}
\caption{9-azidophenanthrene molecules on Ag(111) sorted into 6 groups using hand-selected exemplars.  80\x80 nm, 1024\x1024 pixels.}
\label{APT-044-counted}
\end{figure}

\begin{figure}
\centering
\includegraphics[keepaspectratio=true, width=1\linewidth]{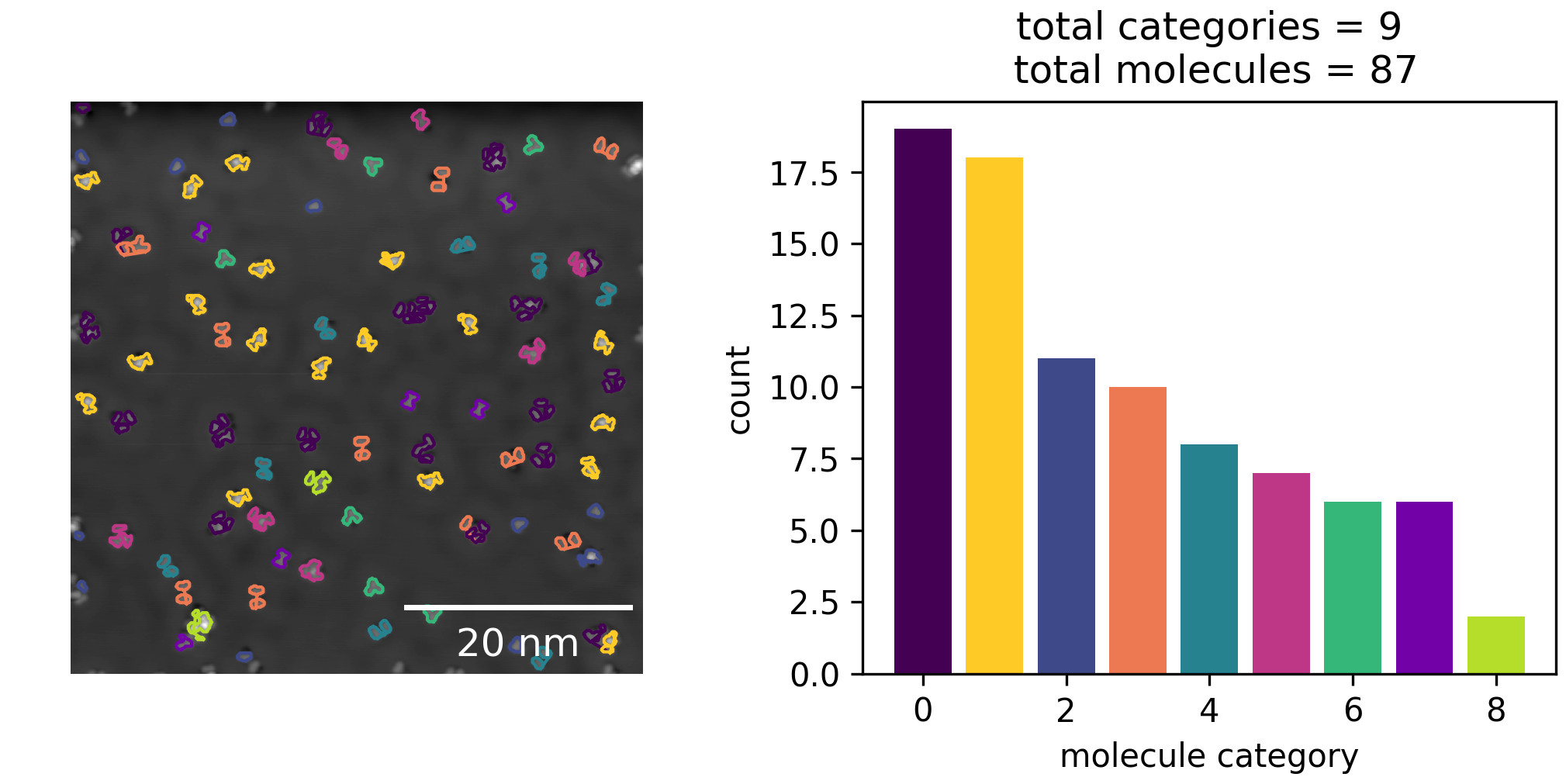}
\caption{9-azidophenanthrene molecules on Ag(111) sorted into 9 groups using hand-selected exemplars.  50\x50 nm, 512\x512 pixels.}
\label{APT-111-counted}
\end{figure}

\begin{figure}
\centering
\includegraphics[keepaspectratio=true, width=1\linewidth]{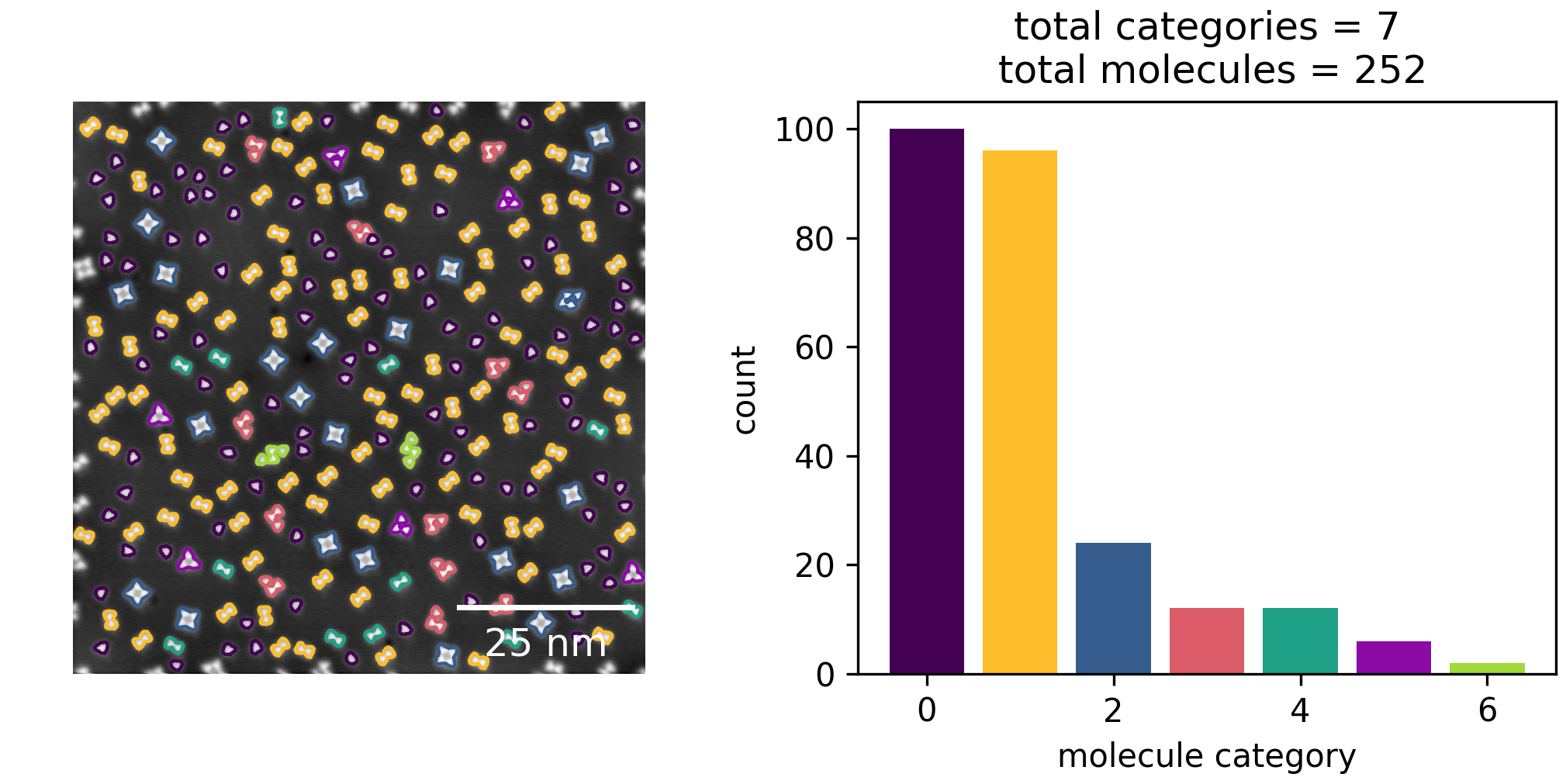}
\caption{Helicene data sorted using hand-selected exemplars.  The dimers (categories 1 and 4) were subsequently sorted by chirality. 80\x80 nm, 1024\x1024 pixels.  Compare to Fig. 4 of Stetsovych \emph{et. al}. \cite{Stetsovych2016}.}
\label{helicene-counted}
\end{figure}

Representative outputs for the three example datasets from hand-selected exemplar molecules are shown in Figs. \ref{APT-044-counted}, \ref{APT-111-counted} and \ref{helicene-counted}.  The work on the 9-azidophenanthrene system which was the source of the data in Figs. \ref{APT-044-counted} and \ref{APT-111-counted} provided the original impetus to develop this package \cite{Hellerstedt2019}.  The helicene data shown in Fig. \ref{helicene-counted} has been previously described in the literature \cite{Stetsovych2016}.  These data were acquired via scanning tunneling microscopy at low temperature ($\sim$ 5~K), with a pixel density of 10 nm$^{-1}$ to 13 nm$^{-1}$.

\section{Expected Impact}

This is the first effort to our knowledge to create an open source scheme to automate the task of counting and sorting hundreds of molecules, specifically tailored for scanning probe microscopy datasets.  We expect this package to be immediately useful for anyone tasked with extracting population statistics from images comparable to those included as examples.  This tool has already proven useful by making it easy to quickly extract and visualize population statistics in image datasets.  By scaling the statistics and quantitative information that can be extracted from these types of datasets, we anticipate that this will facilitate the ability to answer novel and more nuanced questions about how chemical processes unfold on a surface.  We hope to reveal and address limitations by having this set of functions applied to the increasingly diverse ecosystem of on-surface chemistry being explored by the greater scanning probe community.

\section{Conclusions}

Here we have presented our efforts to automate the counting and sorting of molecules in topographic images acquired with a scanning tunneling microscope.  The example datasets, images with several hundred molecules, can be sorted on a personal computer in seconds.  The python based, open source, modular design of the components strives to be immediately accessible to a non-expert user, and allow for significant modification/ customization to suit individual needs.  We hope that use and feedback from the wider community will allow this tool to continue to be developed and provide utility for advancing the understanding of on-surface chemistry experiments.

\section{Software/ Data availability}

\noindent This package is available on GitHub and archived at:\newline
\url{https://doi.org/10.5281/zenodo.6324850}\newline
The data in Figures \ref{APT-044-counted}, \ref{APT-111-counted}, \ref{helicene-counted} is available on figshare: \newline
\url{https://doi.org/10.6084/m9.figshare.19217556}

\section*{Acknowledgements}

S. Edalatmanesh and A. Caparr\'{o}s provided useful discussion in developing the solution to the chirality sorting problem.  We thank Dr. Karth{\"a}user for critical reading of this manuscript. J.H. acknowledges the support of the CIES/ Czech Fulbright commission during the development of this package.



\bibliographystyle{elsarticle-num} 
\bibliography{Mendeley.bib}

\begin{thebibliography}{10}
\expandafter\ifx\csname url\endcsname\relax
  \def\url#1{\texttt{#1}}\fi
\expandafter\ifx\csname urlprefix\endcsname\relax\def\urlprefix{URL }\fi
\expandafter\ifx\csname href\endcsname\relax
  \def\href#1#2{#2} \def\path#1{#1}\fi

\bibitem{Barth2007}
J.~V. Barth,
  \href{http://www.annualreviews.org/doi/abs/10.1146/annurev.physchem.56.092503.141259}{{Molecular
  Architectonic on Metal Surfaces}}, Annual Review of Physical Chemistry 58~(1)
  (2007) 375--407.
\newblock \href {http://dx.doi.org/10.1146/annurev.physchem.56.092503.141259}
  {\path{doi:10.1146/annurev.physchem.56.092503.141259}}.
\newline\urlprefix\url{http://www.annualreviews.org/doi/abs/10.1146/annurev.physchem.56.092503.141259}

\bibitem{Jelinek2017}
P.~Jel{\'{i}}nek,
  \href{http://stacks.iop.org/0953-8984/29/i=34/a=343002?key=crossref.eee0848de87e9b5c6b61380d39a7cee4}{{High
  resolution SPM imaging of organic molecules with functionalized tips}},
  Journal of Physics: Condensed Matter 29~(34) (2017) 343002.
\newblock \href {http://dx.doi.org/10.1088/1361-648X/aa76c7}
  {\path{doi:10.1088/1361-648X/aa76c7}}.
\newline\urlprefix\url{http://stacks.iop.org/0953-8984/29/i=34/a=343002?key=crossref.eee0848de87e9b5c6b61380d39a7cee4}

\bibitem{Capsoni2016}
M.~C. Capsoni,
  \href{https://open.library.ubc.ca/cIRcle/collections/ubctheses/24/items/1.0305019}{{On-surface
  self-assembly and characterization of a macromolecular charge transfer
  complex by scanning tunneling microscopy and spectroscopy}}, Ph.D. thesis,
  University of British Columbia (2016).
\newblock \href {http://dx.doi.org/10.14288/1.0305019}
  {\path{doi:10.14288/1.0305019}}.
\newline\urlprefix\url{https://open.library.ubc.ca/cIRcle/collections/ubctheses/24/items/1.0305019}

\bibitem{Prinz2015}
J.~Prinz, O.~Gr{\"{o}}ning, H.~Brune, R.~Widmer, {Highly enantioselective
  adsorption of small prochiral molecules on a chiral intermetallic compound},
  Angewandte Chemie - International Edition 54~(13) (2015) 3902--3906.
\newblock \href {http://dx.doi.org/10.1002/anie.201410107}
  {\path{doi:10.1002/anie.201410107}}.

\bibitem{Stetsovych2016}
O.~Stetsovych, M.~{\v{S}}vec, J.~Vacek, J.~V. Chocholou{\v{s}}ov{\'{a}},
  A.~Jan{\v{c}}a{\v{r}}{\'{i}}k, J.~Ryb{\'{a}}{\v{c}}ek, K.~Kosmider, I.~G.
  Star{\'{a}}, P.~Jel{\'{i}}nek, I.~Star{\'{y}},
  \href{http://www.nature.com/doifinder/10.1038/nchem.2662}{{From helical to
  planar chirality by on-surface chemistry}}, Nature Chemistry 9~(3) (2016)
  213--218.
\newblock \href {http://dx.doi.org/10.1038/nchem.2662}
  {\path{doi:10.1038/nchem.2662}}.
\newline\urlprefix\url{http://www.nature.com/doifinder/10.1038/nchem.2662}

\bibitem{Goll2022}
F.~D. Goll, G.~Taubmann, U.~Ziener,
  \href{https://onlinelibrary.wiley.com/doi/10.1002/anie.202117580}{{Static
  Scanning Tunneling Microscopy Images Reveal the Mechanism of Supramolecular
  Polymerization of an Oligopyridine on Graphite}}, Angewandte Chemie
  International Edition\href {http://dx.doi.org/10.1002/anie.202117580}
  {\path{doi:10.1002/anie.202117580}}.
\newline\urlprefix\url{https://onlinelibrary.wiley.com/doi/10.1002/anie.202117580}

\bibitem{Horcas2007}
I.~Horcas, R.~Fernández, J.~M. Gómez-Rodríguez, J.~Colchero,
  J.~Gómez-Herrero, A.~M. Baro,
  \href{http://scitation.aip.org/content/aip/journal/rsi/78/1/10.1063/1.2432410}{{WSXM:
  A software for scanning probe microscopy and a tool for nanotechnology}},
  Review of Scientific Instruments 78~(1) (2007) 013705.
\newblock \href {http://dx.doi.org/10.1063/1.2432410}
  {\path{doi:10.1063/1.2432410}}.
\newline\urlprefix\url{http://scitation.aip.org/content/aip/journal/rsi/78/1/10.1063/1.2432410}

\bibitem{Necas2012}
D.~Ne{\v{c}}as, P.~Klapetek, {Gwyddion: An open-source software for SPM data
  analysis}, Central European Journal of Physics 10~(1) (2012) 181--188.
\newblock \href {http://dx.doi.org/10.2478/s11534-011-0096-2}
  {\path{doi:10.2478/s11534-011-0096-2}}.

\bibitem{Rueden2017}
C.~T. Rueden, J.~Schindelin, M.~C. Hiner, B.~E. DeZonia, A.~E. Walter, E.~T.
  Arena, K.~W. Eliceiri, {ImageJ2: ImageJ for the next generation of scientific
  image data}, BMC Bioinformatics 18~(1) (2017) 1--26.
\newblock \href {http://dx.doi.org/10.1186/s12859-017-1934-z}
  {\path{doi:10.1186/s12859-017-1934-z}}.

\bibitem{Cognard2020}
M.~Cognard, \href{https://perma.cc/4K5V-JQRE
  https://www.digitalsurf.com/news/perform-a-particle-analysis-on-microscopy-images/}{{Digital
  Surf}} (2020).
\newline\urlprefix\url{https://perma.cc/4K5V-JQRE
  https://www.digitalsurf.com/news/perform-a-particle-analysis-on-microscopy-images/}

\bibitem{Scherbela2017}
M.~Scherbela, L.~H{\"{o}}rmann, A.~Jeindl, V.~Obersteiner, O.~T. Hofmann,
  \href{http://arxiv.org/abs/1709.05417
  https://link.aps.org/doi/10.1103/PhysRevMaterials.2.043803}{{Charting the
  energy landscape of metal/organic interfaces via machine learning}}, Physical
  Review Materials 2~(4) (2018) 043803.
\newblock \href {http://dx.doi.org/10.1103/PhysRevMaterials.2.043803}
  {\path{doi:10.1103/PhysRevMaterials.2.043803}}.
\newline\urlprefix\url{http://arxiv.org/abs/1709.05417
  https://link.aps.org/doi/10.1103/PhysRevMaterials.2.043803}

\bibitem{Ziatdinov2017}
M.~Ziatdinov, O.~Dyck, A.~Maksov, X.~Li, X.~Sang, K.~Xiao, R.~R. Unocic,
  R.~Vasudevan, S.~Jesse, S.~V. Kalinin,
  \href{http://pubs.acs.org/doi/10.1021/acsnano.7b07504}{{Deep Learning of
  Atomically Resolved Scanning Transmission Electron Microscopy Images:
  Chemical Identification and Tracking Local Transformations}}, ACS Nano
  11~(12) (2017) 12742--12752.
\newblock \href {http://dx.doi.org/10.1021/acsnano.7b07504}
  {\path{doi:10.1021/acsnano.7b07504}}.
\newline\urlprefix\url{http://pubs.acs.org/doi/10.1021/acsnano.7b07504}

\bibitem{Li2021}
J.~Li, M.~Telychko, J.~Yin, Y.~Zhu, G.~Li, S.~Song, H.~Yang, J.~Li, J.~Wu,
  J.~Lu, X.~Wang, \href{https://pubs.acs.org/doi/10.1021/jacs.1c03091}{{Machine
  Vision Automated Chiral Molecule Detection and Classification in Molecular
  Imaging}}, Journal of the American Chemical Society 143~(27) (2021)
  10177--10188.
\newblock \href {http://dx.doi.org/10.1021/jacs.1c03091}
  {\path{doi:10.1021/jacs.1c03091}}.
\newline\urlprefix\url{https://pubs.acs.org/doi/10.1021/jacs.1c03091}

\bibitem{Rashidi2018}
M.~Rashidi, R.~A. Wolkow, \href{http://arxiv.org/abs/1803.07059
  http://pubs.acs.org/doi/10.1021/acsnano.8b02208}{{Autonomous Scanning Probe
  Microscopy in Situ Tip Conditioning through Machine Learning}}, ACS Nano
  (2018) acsnano.8b02208\href {http://dx.doi.org/10.1021/acsnano.8b02208}
  {\path{doi:10.1021/acsnano.8b02208}}.
\newline\urlprefix\url{http://arxiv.org/abs/1803.07059
  http://pubs.acs.org/doi/10.1021/acsnano.8b02208}

\bibitem{Krull2020}
A.~Krull, P.~Hirsch, C.~Rother, A.~Schiffrin, C.~Krull,
  \href{http://dx.doi.org/10.1038/s42005-020-0317-3
  http://www.nature.com/articles/s42005-020-0317-3}{{Artificial-intelligence-driven
  scanning probe microscopy}}, Communications Physics 3~(1) (2020) 54.
\newblock \href {http://dx.doi.org/10.1038/s42005-020-0317-3}
  {\path{doi:10.1038/s42005-020-0317-3}}.
\newline\urlprefix\url{http://dx.doi.org/10.1038/s42005-020-0317-3
  http://www.nature.com/articles/s42005-020-0317-3}

\bibitem{Hellerstedt2019}
J.~Hellerstedt, A.~Cahl{\'{i}}k, O.~Stetsovych, M.~{\v{S}}vec, T.~K. Shimizu,
  P.~Mutombo, J.~Kl{\'{i}}var, I.~G. Star{\'{a}}, P.~Jel{\'{i}}nek,
  I.~Star{\'{y}}, \href{http://doi.wiley.com/10.1002/anie.201812334}{{Aromatic
  Azide Transformation on the Ag(111) Surface Studied by Scanning Probe
  Microscopy}}, Angewandte Chemie International Edition 58~(8) (2019)
  2266--2271.
\newblock \href {http://dx.doi.org/10.1002/anie.201812334}
  {\path{doi:10.1002/anie.201812334}}.
\newline\urlprefix\url{http://doi.wiley.com/10.1002/anie.201812334}

\bibitem{Khotanzad1990}
A.~Khotanzad, Y.~H. Hong, {Invariant Image Recognition by Zernike Moments},
  Ann. Oper. Res. Pattern Anal. Machine Intell. IEEE Trans. Pattern Anal.
  Machine Intell. J . Robotics Res. J . Robotics Res. J . ACM Networks I . J .
  Stoker 12~(14) (1990) 13--118.
\newblock \href {http://dx.doi.org/10.1109/34.55109}
  {\path{doi:10.1109/34.55109}}.

\bibitem{Coelho2013}
L.~P. Coelho,
  \href{http://openresearchsoftware.metajnl.com/articles/10.5334/jors.ac/}{{Mahotas:
  Open source software for scriptable computer vision}}, Journal of Open
  Research Software 1~(1) (2013) e3.
\newblock \href {http://dx.doi.org/10.5334/jors.ac}
  {\path{doi:10.5334/jors.ac}}.
\newline\urlprefix\url{http://openresearchsoftware.metajnl.com/articles/10.5334/jors.ac/}

\bibitem{Oliphant2007}
T.~E. Oliphant, \href{http://ieeexplore.ieee.org/document/4160250/}{{Python for
  Scientific Computing}}, Computing in Science {\&} Engineering 9~(3) (2007)
  10--20.
\newblock \href {http://dx.doi.org/10.1109/MCSE.2007.58}
  {\path{doi:10.1109/MCSE.2007.58}}.
\newline\urlprefix\url{http://ieeexplore.ieee.org/document/4160250/}

\bibitem{VanderWalt2014}
S.~van~der Walt, J.~L. Sch{\"{o}}nberger, J.~Nunez-Iglesias, F.~Boulogne, J.~D.
  Warner, N.~Yager, E.~Gouillart, T.~Yu,
  \href{https://peerj.com/articles/453}{{scikit-image: image processing in
  Python}}, PeerJ 2 (2014) e453.
\newblock \href {http://dx.doi.org/10.7717/peerj.453}
  {\path{doi:10.7717/peerj.453}}.
\newline\urlprefix\url{https://peerj.com/articles/453}

\bibitem{Pedregosa2011}
F.~Pedregosa, G.~Varoquaux, A.~Gramfort, V.~Michel, B.~Thirion, O.~Grisel,
  M.~Blondel, P.~Prettenhofer, R.~Weiss, V.~Dubourg, J.~Vanderplas, A.~Passos,
  D.~Cournapeau, M.~Brucher, M.~Perrot, E.~Duchesnay, {Scikit-learn: Machine
  Learning in Python}, Journal of Machine Learning Research 12 (2011)
  2825--2830.

\bibitem{Hunter2007}
J.~D. Hunter, \href{http://ieeexplore.ieee.org/document/4160265/}{{Matplotlib:
  A 2D Graphics Environment}}, Computing in Science {\&} Engineering 9~(3)
  (2007) 90--95.
\newblock \href {http://dx.doi.org/10.1109/MCSE.2007.55}
  {\path{doi:10.1109/MCSE.2007.55}}.
\newline\urlprefix\url{http://ieeexplore.ieee.org/document/4160265/}

\bibitem{Otsu1979}
N.~Otsu, \href{http://ieeexplore.ieee.org/document/4310076/}{{A Threshold
  Selection Method from Gray-Level Histograms}}, IEEE Transactions on Systems,
  Man, and Cybernetics 9~(1) (1979) 62--66.
\newblock \href {http://dx.doi.org/10.1109/TSMC.1979.4310076}
  {\path{doi:10.1109/TSMC.1979.4310076}}.
\newline\urlprefix\url{http://ieeexplore.ieee.org/document/4310076/}

\bibitem{Zhang1996}
T.~Zhang, R.~Ramakrishnan, M.~Livny,
  \href{http://portal.acm.org/citation.cfm?doid=235968.233324}{{BIRCH: an
  efficient data clustering method for very large databases}}, ACM SIGMOD
  Record 25~(2) (1996) 103--114.
\newblock \href {http://dx.doi.org/10.1145/235968.233324}
  {\path{doi:10.1145/235968.233324}}.
\newline\urlprefix\url{http://portal.acm.org/citation.cfm?doid=235968.233324}

\bibitem{Buda1992}
A.~B. Buda, T.~A. der Heyde, K.~Mislow,
  \href{http://doi.wiley.com/10.1002/anie.199209891}{{On Quantifying
  Chirality}}, Angewandte Chemie International Edition in English 31~(8) (1992)
  989--1007.
\newblock \href {http://dx.doi.org/10.1002/anie.199209891}
  {\path{doi:10.1002/anie.199209891}}.
\newline\urlprefix\url{http://doi.wiley.com/10.1002/anie.199209891}

\end{thebibliography}

\end{document}